\documentclass[11pt]{article}
\usepackage[final]{graphics}

\def\line#1{\hbox to \textwidth{#1}}
\catcode`\@=11

\def\thebibliography#1{\section*{REFERENCES}\list{\arabic{enumi}.}
  {\settowidth\labelwidth{#1.}\leftmargin=1.67em
   \labelsep\leftmargin \advance\labelsep-\labelwidth
   \itemsep\z@ \parsep\z@
   \usecounter{enumi}}\def\makelabel##1{\rlap{##1}\hss}%
   \def\newblock{\hskip 0.11em plus 0.33em minus -0.07em}
   \sloppy \clubpenalty=4000 \widowpenalty=4000 \sfcode`\.=1000\relax}

\def\@cite#1#2{$[{{#1\if@tempswa , #2\fi}}]$}

\newcount\@tempcntc
\def\@citex[#1]#2{\if@filesw\immediate\write\@auxout{\string\citation{#2}}\fi
  \@tempcnta\z@\@tempcntb\m@ne\def\@citea{}\@cite{%
	\@ordonner{#2}%
	\@for\@citeb:=#2\do%
    {\@ifundefined{b@\@citeb}%
	{\@citeo\@tempcntb\m@ne\@citea%
        	\def\@citea{,\penalty\@m\ }{\bf ?}\@warning%
       		{Citation `\@citeb' on page \thepage \space undefined}}%
    	{\setbox\z@\hbox{\global\@tempcntc0\csname b@\@citeb\endcsname\relax}
     \ifnum\@tempcntc=\z@ \@citeo\@tempcntb\m@ne%
       \@citea\def\@citea{,\penalty\@m}%
       \hbox{\csname b@\@citeb\endcsname}%
     \else%
      \advance\@tempcntb\@ne%
      \ifnum\@tempcntb=\@tempcntc%
      \else\advance\@tempcntb\m@ne\@citeo%
      \@tempcnta\@tempcntc\@tempcntb\@tempcntc\fi\fi}}\@citeo}{#1}}%

\def\@citeo{\ifnum\@tempcnta>\@tempcntb\else\@citea
  \def\@citea{,\penalty\@m}%
  \ifnum\@tempcnta=\@tempcntb\the\@tempcnta\else
   {\advance\@tempcnta\@ne\ifnum\@tempcnta=\@tempcntb \else
\def\@citea{-}\fi
    \advance\@tempcnta\m@ne\the\@tempcnta\@citea\the\@tempcntb}\fi\fi}

\def\@toto{}
\newif\if@ordre 
\newcount\c@current
\newcount\c@last

\def\@ordonner#1{\global\c@last\m@ne%
		\global\@ordretrue%
		\@for\@toto:=#1\do%
			{\@ifundefined{b@\@toto}%
			{}%
			{\c@current\csname b@\@toto\endcsname\relax%
			\ifnum\the\c@current<\the\c@last\relax%
				{\global\@ordrefalse}\fi%
			\global\c@last\the\c@current%
			}%
			}%
		\if@ordre{}\else{\typeout{}%
			\typeout{Warning: the references are not %
			 in increasing order\on@line:}%
			\@for\@toto:=#1\do%
			{\@ifundefined{b@\@toto}%
			{}%
			\typeout{\@toto:\space \@nameuse{b@\@toto}}%
			}\typeout{}}\fi%
		}%

\catcode`\@=12

\newcommand{\slL}{\raise.15ex\hbox{$/$}\kern-.53em\hbox{$L$}}
\newcommand{\slP}{\raise.15ex\hbox{$/$}\kern-.53em\hbox{$P$}}
\newcommand{\slR}{\raise.15ex\hbox{$/$}\kern-.53em\hbox{$R$}}
\newcommand{\slQ}{\raise.15ex\hbox{$/$}\kern-.53em\hbox{$Q$}}
\newcommand{\slK}{\raise.15ex\hbox{$/$}\kern-.53em\hbox{$K$}}

\newcommand{\be}{\begin{equation}}
\newcommand{\ee}{\end{equation}}     
\newcommand{\bea}{\begin{eqnarray}}
\newcommand{\ena}{\end{eqnarray}}

\def\build#1\over#2{\mathrel{\mathop{\kern 0pt#2}\limits_{#1}}}

\font\tenimbf=cmmib10 at 12pt
\font\sevenimbf=cmmib10 at 7pt
\font\fiveimbf=cmmib10 at 5pt
\newfam\imbf
\textfont\imbf=\tenimbf
\scriptfont\imbf=\sevenimbf
\scriptscriptfont\imbf=\fiveimbf
\def\imb{\fam\imbf\tenimbf}

\title{Soft photons and hard thermal loops\footnote{Talk given at the RHIC
Summer Study, 6-16 July 1997, Brookhaven National Laboratory, USA.}}
\author{Fran\c cois Gelis\footnote{E-mail: gelis@lapp.in2p3.fr}\\
\\
Laboratoire de Physique Th\'eorique ENSLAPP\\
B.P. 110, F-74941 Annecy-le-Vieux Cedex, France}

\begin{document}
\bibliographystyle{unsrt}
\maketitle

\begin{abstract}
We calculate within the Hard Thermal Loop expansion the production rate of
soft photons and lepton pairs by a hot quark-gluon plasma at thermal
equilibrium, up to the 2-loop order. Strong collinear divergences appear to
mix the orders of the perturbative expansion so that the 2-loop order is in
fact dominant compared to the 1-loop order. 
More precisely, angular integrals that would otherwise be 
of order 1 are found to
behave as powers of $1/g$ when the produced photon is massless, therefore 
breaking
the standard HTL power counting rules.

\end{abstract}
\line{\hfill ENSLAPP-A-665, hep-ph/9710203 \hfill}

\section{Introduction}
   The production rate of photons or dileptons is of some relevance to
   the phenomenology of a quark-gluon plasma. Indeed, these electromagnetic
   probes are interacting very slightly through electromagnetic forces.
   Therefore, it is expected that they will escape the plasma without
   re-interacting after their creation. As a consequence, they are rather clean
   probes of the state of the system at the time it produced them. 
   
   Among the theoretical tools to calculate such observables, one can quote the
   semi-classical methods based on an eikonal approximation 
   \cite{CleymGR1,CleymGR2,BaierDMPS1,BaierDMPS2}, and the tools of
   finite temperature field theory
   \cite{BraatPY1,BaierPS1,AurenBP1,Wong1,AurenGKP1,AurenGKP2}. 
   In this report, I focus mainly on
   the field theoretical method. I assume a plasma in thermal equilibrium and 
   sufficiently hot so that
   the temperature is much larger than the masses of the particles.
   From the theoretical point of view, studying these production rates is a
   good test of our present understanding of problems such as infrared and
   collinear singularities in high temperature QCD.
   
   The starting point is to connect the photon production rate to the imaginary
   part of the photon polarization tensor, via the following two relations
   \cite{BraatPY1,Weldo3}:
   \begin{eqnarray}
   &&{\rm At\ }Q^2=0{\rm ,}\qquad {{dN}\over{dtd^3{\imb x}}} = - {{d^3{\imb
   q}}\over{(2\pi)^3}}\; {{n_{_{B}}(q_o)}\over{q_o}} {\rm
   Im}\,\Pi^\mu{}_\mu(q_o,{\imb q}) \\
   &&{\rm At\ }Q^2>0{\rm ,}\qquad {{dN}\over{dtd^3{\imb x}}} = - \alpha
   {{dq_od^3{\imb
   q}}\over{12\pi^3}}\; {{n_{_{B}}(q_o)}\over{Q^2}} {\rm
   Im}\,\Pi^\mu{}_\mu(q_o,{\imb q})\; .
   \end{eqnarray}
   Basically, these two formulae differ only by the allowed phase space, 
   by the coupling constant involved
   in the decay of a virtual photon into a lepton pair and by the propagator
   of such a heavy photon\footnote{The above two formulae are true to all
   orders in the strong coupling constant $\alpha_{_{S}}$, but only to the
   first non vanishing order in the QED coupling $\alpha$, due to the
   fact that they neglect potential re-interactions of the produced
   photon on its way out of the plasma.}.

\section{Infrared problems and hard thermal loops}
   \subsection{Origin of the problem}
   Before going on with some details of the calculation we performed, it is
   worth recalling some of the issues related to the concept of hard thermal
   loops (HTL) \cite{BraatP1}, 
   and its relevance to the problem of infrared divergences.
   Indeed, although HTLs have been introduced to solve the problem of the
   apparent non gauge invariance of the gluon damping rate \cite{BraatP3}, 
   they proved to be
   quite efficient in improving the infrared behavior of thermal gauge
   theories, which is {\it a priori} more severe than at $T=0$ due to the
   singular behavior of Bose-Einstein's factors at small energies:
   \begin{equation}
   n_{_{B}}(l_o)={1\over{e^{l_0/T}-1}}\approx {T\over{l_o}} \gg 1 \quad{\rm if\
   } l_o\ll T\; .
   \end{equation}
   Since in the Feynman rules, these factors are always accompanied by the 
   Dirac's distribution $\delta(l_o^2-{\imb l}^2-m^2)$, the Bose-Einstein's
   weights are a problem only if there are massless bosons in the theory. In
   QCD, one has gluons...
   
   \subsection{Semantics: hard and soft scales}
   \begin{figure}[htbp]
   \parbox{5cm}{\caption{
   Hard and soft scales.\label{scales}}}
   \parbox{6cm}{\hfill
   \resizebox*{!}{2cm}{\includegraphics{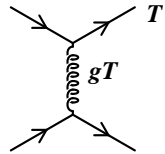}}\hglue 1cm}
   \end{figure} 
   It is useful in thermal field theories to distinguish between two typical
   energy scales: particles of energy of order $T$ are said to be hard, and
   particles of energy of order $gT$ are said to be soft. Roughly speaking, the
   hard scale is the typical energy of partons in the plasma, and the soft scale
   is the typical energy of quanta exchanged during parton interactions.

   \subsection{Summation of hard thermal loops}
   The foundations of the HTL concept lie in the fact that some one-loop
   diagrams have a thermal contribution which is dominated by the hard region of
   phase space, and moreover this thermal contribution is as large as the bare
   corresponding function when all its external legs carry soft momenta. For
   instance, this is what happens for the one-loop photon self-energy in QED.
   \begin{figure}[htbp]
   \parbox{5cm}{\caption{
   Hard ther\-mal loop of the pho\-ton self e\-ner\-gy.\label{gammaHTL}}}
   \parbox{7cm}{\hfill
   \resizebox*{!}{2cm}{\includegraphics{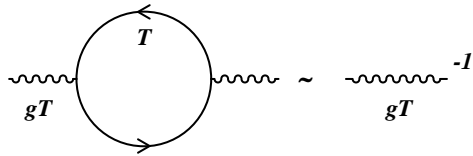}}\hglue 1cm}
   \end{figure}
   Then, in order to take these large contributions into account, one defines
   an effective field theory \cite{BraatP4,FrenkT2} by the means of 
   effective propagators and
   vertices. The effective propagators are obtained by a Dyson summation of the
   HTL contribution to the corresponding self energy, whereas the effective
   vertices are defined to be the sum of their bare counterpart plus the HTL
   contribution to the corresponding vertex function. In order to avoid double
   counting of thermal corrections, one should include counter-terms whose
   purpose is to subtract at higher orders what had been added by this
   summation. In that way, the overall Lagrangian function remains 
   unchanged, so
   that the use of this effective theory just amounts to a re-ordering of the
   perturbative expansion.
   
   \subsection{Basic properties}
   The HTLs have a few nice properties that make the effective
   theories obtained by they summation rather attractive:
   
   \noindent $\bullet$ The HTLs are gauge invariant. As a consequence, the
   effective theory one obtains by summing all the HTLs is also gauge
   invariant. It is important to note the fact that the effective vertices are
   essential here to achieve the gauge invariance of the effective Lagrangian.
   In fact, starting by the summation of 2-point HTLs in order to obtain a
   better behaved effective propagator, one can obtain all the $n-$point 
   $(n\ge
   3)$ HTLs by the requirement that the effective theory should be gauge
   invariant \cite{BraatP4}.
   
   \noindent $\bullet$ The common physical interpretation of the HTLs is that
   they reflect the long distance behavior of the gauge interaction better
   than the bare theory did. In particular, they account for the Debye
   screening phenomenon, which is translated in field theory by the fact 
   that the
   gluon acquires an effective mass $m_{_{D}}\sim gT$, thus leading to a
   screened interaction whose range is of order $m_{_{D}}^{-1}$.
   \begin{figure}[htbp]
   \parbox{5cm}{\caption{
   QED Debye screening.\label{debye}}}
   \parbox{7cm}{\hfill
   \resizebox*{!}{2cm}{\includegraphics{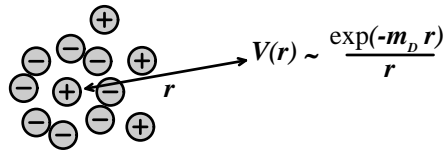}}\hglue 1cm}
   \end{figure}

   \noindent $\bullet$ The $n-$point $(n\ge 3)$ HTLs are vanishing if one takes
   the trace over the Lorentz indices of two external gauge bosons. 
   
   \noindent $\bullet$ Another property which makes the HTLs simple is the fact
   that the 1-loop diagrams that give the HTLs are calculated with massless
   particles running in the loop. Besides the technical simplification this fact
   provides, it is also at the origin of collinear singularities 
   encountered when
   one is using HTLs with light-like external legs.

   \subsection{Further problems}
   Despite the great improvement the HTLs provide for perturbative calculations
   in thermal gauge theories, some nasty problems remain.
   
   \noindent $\bullet$ The first one is the problem of collinear singularities
   evocated above. Due to the fact that the HTLs are calculated with massless
   propagators running in the hard loop, these functions are divergent when
   at least one of the external legs is on-shell. This is due to an angular
   integral like\footnote{This example displays only a logarithmic
   divergence. Nevertheless, when two external momenta are on-shell and
   collinear, one obtains stronger collinear singularities.}:
   \begin{equation}
   \int{{d\Omega_{\imb p}}\over{\widehat{P}\cdot K}} 
   \propto \int\limits_{-1}^{+1}
   {{d\cos \theta}\over{k(1-\cos\theta)}} = \infty\; ,
   \end{equation}
   where $\widehat{P}\equiv(1,\hat{\imb p})$, and $K$ is a light-like external
   momentum. A possible solution to this problem has been proposed recently
   \cite{FlechR1},
   which amounts to keep an asymptotic thermal mass for the particle running in
   the hard loop. Obviously, the previous divergent integral becomes now:
   \begin{equation}
   \int{{d\Omega_{\imb p}}\over{\widehat{P}\cdot K}} 
   \propto \int\limits_{-1}^{+1}
   {{d\cos\theta}\over{\omega_k-k\cos\theta}}< \infty\; ,
   \end{equation}
   where $\omega_k\equiv\surd({\imb k}^2+m^2)$. 
   Moreover, it has been shown that
   these improved hard thermal loops still generate a gauge invariant effective
   theory.
   
   \noindent $\bullet$ The other problem which is not solved by the summation of
   HTLs is due to the fact that static transverse gauge bosons don't get 
   a thermal
   mass. This property is known to be true to all orders in QED, in which case
   it is related to the fact that static magnetic fields are not screened in a
   plasma. In QCD, there is still a hope that self-interactions of gluons
   can generate such a mass for the transverse modes, but it seems that this
   mass will be beyond the abilities of perturbative methods, and at most of
   order $g^2T$ contrary to usual thermal masses that are of order $gT$.
   The nullity of this {\it magnetic mass} is at the origin of further infrared
   divergences, like those encountered in the perturbative 
   calculation of the damping rate of
   a fast fermion.
   
\section{$\gamma$ production rate at $1$ loop}
   \subsection{1-loop diagrams and results}
   	\begin{figure}[h!]  
            \centerline{
            \hfill
                \resizebox*{!}{2.2cm}{\includegraphics{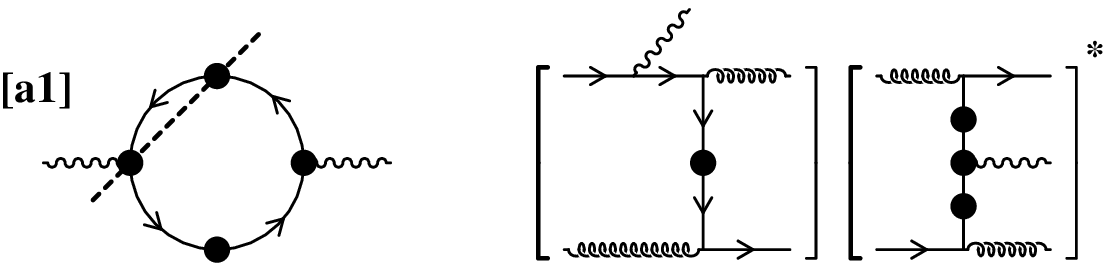}}
                \hglue 0.5cm}
             \vskip 3mm
            \parbox{5cm}{\caption{\footnotesize{}Contributions to the photon production
            rate at 1-loop in the effective theory. The first one 
            is accompanied by the
            corresponding physical amplitudes. Black dots denote effective
            propagators and vertices.}\label{1loop}}
                \parbox{7cm}{
                \hfill\resizebox*{!}{2.2cm}{\includegraphics{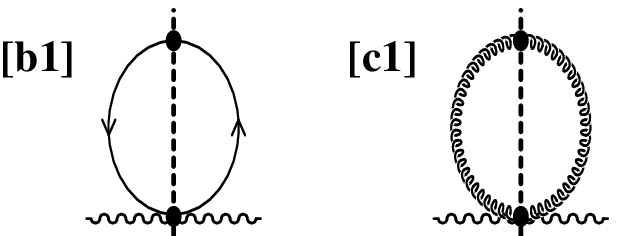}}
                \hglue 5mm}
      \end{figure} 
   
   These 1-loop diagrams have been evaluated by several groups
   \cite{BraatPY1,BaierPS1,AurenBP1,Wong1}, leading to the
   following results:
   \begin{eqnarray}
        &&{\rm Im}\,\Pi^\mu{}_\mu{}_{|{\imb [a1]}}(q_o,{\imb q})
        \sim e^2 g^4 {{T^3}\over{q_o}}\; ,\\
        &&{\rm Im}\,\Pi^\mu{}_\mu{}_{|{\imb [b1]}}(q_o,{\imb q})
        =0\; ,\qquad {\rm Im}\,\Pi^\mu{}_\mu{}_{|{\imb [c1]}}(q_o,{\imb q})
        =0\; .
\end{eqnarray}
The result given for the diagram $[{\imb a1}]$ is to be preceded by a numerical
coefficient which is finite if the emitted photon is massive $(Q^2> 0)$, but
that diverges logarithmically when the photon mass goes to zero. This is due to
the fact that the hard loop contained in the effective $\gamma q \bar{q}$ 
vertex is plagued by a collinear divergence when one of its external legs is
light-like. If one applies the prescription presented in the previous section,
this coefficient becomes a number of order $\ln (1/g)$. 

   \subsection{Comments}
   A few remarks are worth saying concerning these results:

\noindent $\bullet$ The only non-vanishing contribution at 1-loop in the
effective theory based on the summation of the HTLs comes from the diagram 
$[{\imb a1}]$ and corresponds to the interference between the two amplitudes
represented in the figure \ref{1loop}. 
If, instead of using thermal field theory, one had
just tried to guess what are the dominant processes contributing to photon
production by a hot plasma, it is unlikely that these processes would have been
thought as being dominant.

\noindent $\bullet$ The exact nullity of the diagrams $[{\imb b1}]$ and $[{\imb
c1}]$ is a consequence of the remark made earlier concerning the trace over
Lorentz indices of $n-$point HTLs.

\noindent $\bullet$ In the case of the diagram $[{\imb a1}]$, taking the trace
$\cdots{}^\mu{}_\mu$ doesn't make the result vanish, but nevertheless is at the
origin of a suppression by a power of $g$. 
Therefore, if this suppression is
specific to this topology, it may be that some 2-loop diagrams are of the same
order (in thermal field theories the expansion parameter is $g$ instead of
$g^2$). 

This fact, added to the oddity of the dominant physical process at 1-loop,
should incite one to evaluate the 2-loop contributions.

\section{$\gamma$ production rate at $2$ loops}
   \subsection{2-loop diagrams}
   In the list of 2-loop diagrams contributing to the photon production 
   rate, we
   skipped right from the beginning those having a {\it tadpole} topology since
   taking the $\cdots{}^\mu{}_\mu$ trace will make them exactly cancel.
   Moreover, we skipped topologies containing effective vertices like $\gamma g
   g g $ or $\gamma g q \bar{q}$. This will be justified later by the fact that
   the topologies $[{\imb a2}]$ and $[{\imb b2}]$ are important because of 
   quasi
   superposed collinear singularities (leading to an enhancement by powers of
   $1/g$), whereas the diagrams we skipped can have at most powers of
   $\ln(1/g)$. For the remaining diagrams, it is {\it a priori} essential to
   keep the topologies with counter-terms in order to avoid double counting of
   thermal corrections already included at the 1-loop level. 
   Nevertheless, it is obvious that the diagrams with
   counter-terms are of the order of magnitude of $[{\imb a1}]$\footnote{If the
   effective theory works properly, the diagrams $[{\imb a2}]$ 
   and $[{\imb b2}]$
   should be partly compensated by these counter-term diagrams, leaving 
   us with a
   sub-dominant correction.} (the counter-term
   is part of the effective vertex contained in $[{\imb a1}]$). 
   \begin{figure}[htbp]  
            \centerline{\parbox{35mm}{\caption{\footnotesize{}Some 2-loop 
            contributions to photon
            production. The cross denote HTL counter-terms.
            \label{2loops}}}\hfill
              \parbox{81mm}{
              \resizebox*{!}{2.1cm}{\includegraphics{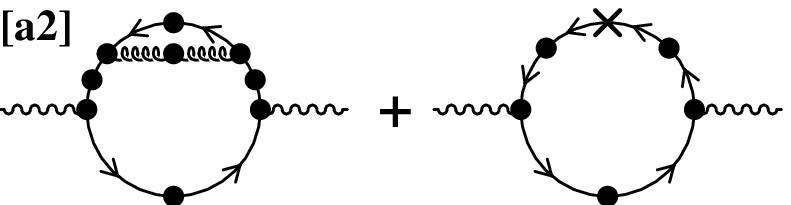}}}}
            \vskip 2mm
            \centerline{\hfill
                \resizebox*{!}{2.1cm}{\includegraphics{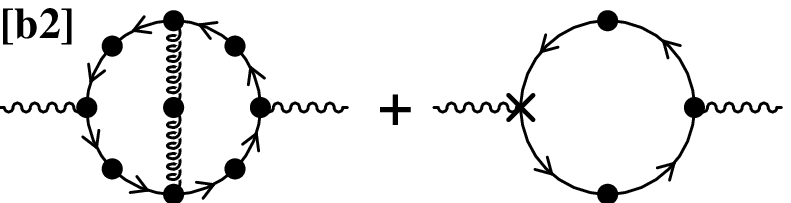}}}
        \end{figure} 

   \subsection{Physical processes}
\begin{figure}[htbp]  
            \parbox{3.5cm}{\caption{\footnotesize{}Simplified version of $\imb [a2]$
        and $\imb [b2]$.\label{2loopsimple}}}
              \parbox{8.7cm}{\hglue 3mm
              \resizebox*{!}{2.3cm}{\includegraphics{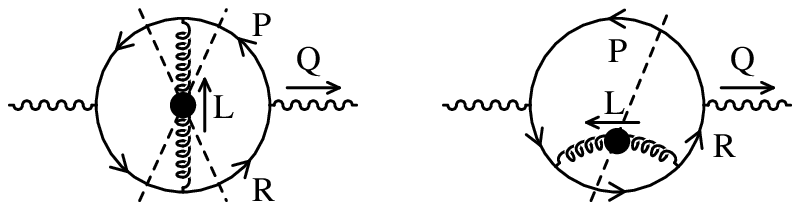}}}
        \end{figure}

   Looking more carefully at the kinematics of these 2-loop diagrams, we can see
   that the dominant contribution comes from the region of phase space where the
   quark running in the loop is hard while the exchanged gluon is soft. As a
   consequence, we can approximate the effective vertices by their bare
   counterpart, and the effective quark propagators by their hard approximation:
   \begin{equation}
   {\cal S}(P)=i{{p_o\gamma^o-\omega_{+}(p)\hat{\imb p}\cdot{\bf\gamma}}
   \over{p_o^2-\omega_{+}^2(p)+i\epsilon}}\; ,
   \end{equation}
   with $\omega_+(p)\equiv\surd({\imb p}^2+M_{_{F}}^2)$, where
   $M^2_{_{F}}\equiv g^2C_{_{F}}T^2/4$.
   Therefore, we can replace the diagrams $[{\imb a2}]$ and $[{\imb b2}]$ by the
   simpler version of the figure \ref{2loopsimple}, where we used a slightly
   abusive notation for the quark propagators we represented as bare ones
   whereas we are retaining an asymptotic thermal mass in them. 
        On the same figure,
   we also represented the cuts relevant to the calculation of the imaginary
   part of the polarization tensor. The cuts not represented always imply that
   the quark be soft, which would drastically reduce the phase space available.
   {\it A priori}, by cutting an effective gluon propagator, we can pick the
   pole part $(L^2>0)$ or the Landau damping part $(L^2<0)$ of the gluon
   spectral function. The two contributions correspond respectively to photon
   production by Compton effect or by bremsstrahlung, which looks more
   intuitive than
   the processes involved in $[{\imb a1}]$.
	\begin{figure}[htbp]  
            \centerline{
            \hfill
                \resizebox*{!}{2cm}{\includegraphics{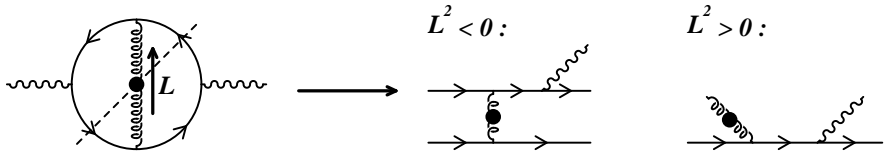}}
                \hglue 0.5cm}
            \caption{\footnotesize{}Physical processes involved 
            in $\imb [b2]$.\label{2loopphys}}
        \end{figure}
        
   \subsection{Collinear behavior}
   If both the external photon and the internal quark are massless, the diagrams
   $[{\imb a2}]$ and $[{\imb b2}]$ exhibits a collinear singularity when the
   3-momentum of the quark is collinear to that of the external photon.
   Moreover, contrary to the case of the diagram $[{\imb a1}]$ where such
   singularities are at most logarithmic, we have here two denominators
   (corresponding to the propagators which are not cut) that vanish almost
   simultaneously. This is obvious in the case of the diagram $[{\imb a2}]$ for
   which we have an exact double pole due to the insertion of the self-energy
   correction. In the case of the diagram $[{\imb b2}]$, both denominators are
   distinct and so are the two poles. At $Q^2=0$, $R^2$ is vanishing 
   when $\imb p$
   is parallel to $\imb q$, and $(P+L)^2$ is vanishing when ${\imb r}+{\imb l}$
   is parallel to $\imb q$. Nevertheless, since $\imb p$ and $\imb r$ are hard
   whereas $\imb q$ and $\imb l$ are soft, these two conditions of collinearity
   are satisfied almost at the same time. It means that the two poles are very
   close, so that their association behaves almost like a double pole.
   
   Therefore, in both $[{\imb a2}]$ and  $[{\imb b2}]$, we have denominators
   that lead to a linear collinear divergence. A careful calculation of the
   angular integral while keeping $Q^2\ge 0$ and the quark thermal mass
   $M_{_{F}}$ indicates that these divergences are regularized by the
   combination $M^2_{\rm eff}\equiv M^2_{_{F}}+Q^2r^2/q_o^2$ divided by $T^2$. 
   If we limit ourselves to photons of very small virtuality $(Q^2\ll q_o^2)$,
   then this dimensionless regulator is much smaller than $1$, and the
   dimensionless angular integral is much larger than $1$, which would be its
   order of magnitude in the absence of collinear problems. In fact, at $Q^2=0$,
   the only scale in the angular integral is $M^2_{_{F}}/T^2\sim g^2$, so that
   the enhancement factor of this integral is $1/g^2$. The fact that powers of
   $1/g$ can appear at $Q^2=0$ in the perturbative expansion 
   due to collinear singularities can make the order of magnitude of a diagram
   somewhat unpredictable by the power counting rules associated usually with
   the HTL expansion (indeed, these rules always assume that the dimensionless
   angular integrals that one can factorize out of the HTLs are of order $1$,
   and we have shown by this example that these integrals can sometimes be as
   large as powers of $1/g$).
   
   The above results concerning the order of magnitude of the angular integrals
   indicate that the result is completely dominated by the collinear sector of
   the phase space. As a consequence, one can use a collinear approximation to
   obtain a simplified expression for the Dirac's algebra associated with the
   quark loop. The results one obtains that way are:
   \begin{equation}
       {\rm Tr}_{\imb [a2]}\approx 16 Q^2 P_\alpha P_\beta\;,\qquad
       {\rm Tr}_{\imb [b2]}\approx 16 (Q^2-L^2) 
       P_\alpha P_\beta \; ,
   \end{equation}
   where $\alpha$ and $\beta$ are the Lorentz indices of the gluon. Then,
   limiting ourselves to very small $Q^2$, we can neglect the
   contribution of the diagram $[{\imb a2}]$.

   Moreover, since we have the constraint $\delta(P^2-M^2_{_{F}})$, the
   denominator $(P+L)^2-M^2_{_{F}}$ can approach zero only if $L^2$ is negative.
   As a consequence, the portion of phase space that contains the collinear
   singularities also verifies $L^2< 0$, which indicates that the bremsstrahlung
   production of photons dominates with respect to the Compton effect.

   \subsection{Results}
\begin{figure}[htbp]
            \parbox{5cm}{\caption{\footnotesize{}Transverse and longitudinal
            contributions as a function of $Q^2/q_o^2$, for $3$ colors and $3$
            light flavors. The value taken for the coupling constant is
            $g=0.44$.\label{plotQ}}}
            \parbox{6cm}{
            \hfill
                \resizebox*{!}{3.5cm}{\includegraphics{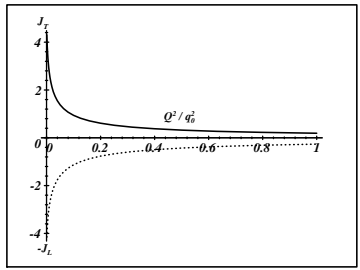}}
                \hglue 5mm}
        \end{figure}

   Therefore, the result we obtain for the imaginary part of the polarization
   tensor of the photon can take the form \cite{AurenGKP1,AurenGKP2}:
      \begin{equation}
        {\rm Im}\,\Pi^\mu{}_\mu(q_o,{\imb q})\approx 
        -{{e^2g^2NC_{_{F}}}\over{3\pi^2}}\sum\limits_{T,L}\,J_{_{T,L}}\,
                {{T^3}\over{q_o}} \; ,
        \end{equation}
        where the quantities $J_{_{T,L}}$ are pure numbers quantifying the
        respective contributions of transverse and longitudinal gluons exchange.
        These coefficients are given by an integral which is evaluated
        numerically, as a function of the ratio $Q^2/q_o^2$ and plotted 
        on the figure
        \ref{plotQ}. We can see clearly
        an enhancement in the region of small $Q^2$. Indeed, since the
        regularization of collinear divergences is realized by the mass
        $M^2_{\rm eff}$, increasing with $Q^2$, the result is larger at small
        $Q^2$. 
        The same quantities are plotted also with a logarithmic scale that
        enables one to see in more detail the region of small $Q^2$. In
        particular, we see that they converge to finite values when $Q^2$ goes
        to $0$. Moreover, we see that the enhancement effect (quantified by the
        ratio of the values taken at $Q^2=0$ and at $Q^2\sim q_o^2$) is larger
        when $g$ is small. 
        \begin{figure}[htbp]
            \parbox{4.8cm}{\caption{\footnotesize{}Effect of the value of the
            coupling constant. Solid lines: $g=0.1$, dashed lines: $g=0.44$,
            dotted lines: $g=2$. We used $3$ colors and $3$ light 
            flavors.\label{plotQg}} }
            \parbox{6.2cm}{
            \hfill
                \resizebox*{!}{3.5cm}{\includegraphics{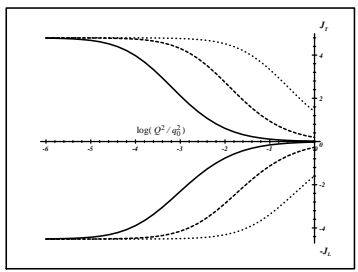}}
                \hglue 5mm}
        \end{figure}
        
        \noindent Let us now list a few 
        important features of the result we obtained:
        
        \noindent $\bullet$ The diagram $[{\imb a1}]$, only non vanishing
        contribution at the 1-loop order, is negligeable compared to $[{\imb
        b2}]$, for small enough $Q^2$ $(Q^2\ll q_o^2 \ll T^2)$. 
        Moreover, the latter is much larger than the diagrams with
        counter-terms of the figure \ref{2loops}. This fact is due to strong
        collinear divergences. Therefore, there is a possibility that such
        collinear singularities may invalidate the
        perturbative expansion since powers of the inverse coupling constant
        emerge in the process.
        
        \noindent $\bullet$ The result we obtained is free of any infrared
        divergence and is totally insensitive to the magnetic mass scale as long
        as the magnetic mass is negligeable in front of the thermal mass
        $M_{_{F}}$. If this condition is satisfied, then the regularization of
        potential IR singularities in the transverse gluon contribution is done
        by the quark mass $M_{_{F}}$. There is a natural explanation for this
        {\it a priori} surprising property related to the fact that the delta
        function
        constraints become incompatible in the limit $L\to 0$ when $M_{_{F}}>0$.
        
        \noindent $\bullet$ Our result can be put into a form that looks like
        the expressions obtained by semi-classical methods:
        \begin{eqnarray}
        {{dN}\over{dtd^3{\imb x}}}\approx&& {{d^3{\imb q}}\over
        {(2\pi)^3 2q_o}} \int
        \prod\limits_{i=1,2}^{}{{d^4P_i}\over{(2\pi)^4}}
        \;2\pi\delta(P_i^2-M_{_{F}}^2)\; n_{_{F}}(p^o_i)\nonumber\\
        &&\times\int  \prod\limits_{i=1,2}^{}
        {{d^4P^\prime_i}\over{(2\pi)^4}}
        \;2\pi\delta(P^\prime_i{}^2-M_{_{F}}^2)\,[1-
        n_{_{F}}(p^\prime_i{}^o)] \nonumber\\ &&\times\left|{\cal
        M} \right|^2(P_1,P_2;P^\prime_1+Q,P^\prime_2) \;\;\;
        (2\pi)^4\delta(P_1+P_2-P^\prime_1-P^\prime_2-Q)\nonumber\\
        &&\times e^2\sum_{{\rm pol.\ }\epsilon}\left(
        {{P_1\cdot\epsilon} \over{P_1\cdot Q}} - {{P^\prime_1\cdot
        \epsilon}\over{P^\prime_1\cdot Q}} \right)^2 \; .
        \label{semi}
        \end{eqnarray}
	The interpretation of the various factors entering the above formula
	is rather easy: starting from the beginning, we find the photon phase
	space, the phase space of two incoming quarks, the phase space of the
	same outgoing quarks, the amplitude $|{\cal M}|^2$ of the scattering
	process between the two quarks, associated to a delta constraint
	ensuring the overall momentum conservation, and finally a factor which
	is the square of the electromagnetic current coupling the photon to the
	quark line. 
	
	Therefore, we see that thermal field theory gives the same
	kind of expression as the one obtained by semi-classical methods
	\cite{CleymGR1,CleymGR2,BaierDMPS1,BaierDMPS2} when
	considering the photon emission induced by a single scattering. In fact,
	the thermal field theory result is more precise in the sense that it has
	not neglected the contribution of transverse gluons, which is as
	important as the longitudinal one as can be seen from the curves plotted
	above.

\section{Conclusions and perspectives}
 	The study of the photon production rate by a hot quark-gluon plasma
	by using the effective theory based on the summation of the hard thermal
	loops has shown that serious problems may arise from strong collinear
	singularities. More precisely, angular integrals that are 
	usually assumed
	to be of order unity in the HTL framework are found to behave as powers
	of $1/g$. 
	It is not known by now if such singularities or even stronger ones may
	emerge from higher order contributions, enabling them to remain at
	the same level.
	
	From a more phenomenological point of view, we arrived at the conclusion
	that real or slightly virtual soft 
	photons are mostly produced by bremsstrahlung in a hot plasma. 
	
	An important issue is to determine whether higher order contributions,
	corresponding to photon emission induced by multiple scatterings, are
	important or not. The importance of this study is related to the fact
	that the semi-classical methods predict that multiple scatterings are
	important and can modify the photon spectrum in the region of small
	energy, a phenomenon known as the Landau-Pomeranchuk-Migdal effect. 

\section*{Acknowledgments}
	It is a pleasure to thank the organizers of this workshop, as well as
	the HET group at BNL for financial support. I must also thank P.
	Aurenche for many useful comments on this report.

\end{document}